\documentclass[11pt]{article}
\usepackage{wspro}
\usepackage{amssymb} 
\usepackage{graphics}
\usepackage{color} 
\makeatletter
\@addtoreset{equation}{section}
\makeatother

\newcommand{\mc}{\mathcal} 

\def\cV{{\mc V}}
\def\cJ{{\mc J}}
\def\cH{{\mc H}}

\def\rmY{{\it Y}}

\def\cR{{\mc R}}
\def\IP{\relax{\rm I\kern-.18em P}}

\newcommand{\nc}{\newcommand}
\newcommand{\beq}{\begin{equation}}
\newcommand{\be}{\begin{equation}}
\newcommand{\eeq}{\end{equation}}
\newcommand{\ee}{\end{equation}}
\newcommand{\beqa}{\begin{eqnarray}}
\newcommand{\ba}{\begin{eqnarray}}
\newcommand{\eeqa}{\end{eqnarray}}
\newcommand{\ea}{\end{eqnarray}}

\def\Mat{{\mbox{\rm Mat}}}

\def\ik{{\sf k}}
\def\min{{\mbox{\rm min\/}}}
\def\cH{{\mc H}}

\def\tr{{\rm tr}}

\def\cC{{\mc C}}

\def\c{\gamma} 
\def\pl{\partial} 
\def\bpl{\bar \partial} 
\def\H3p{H_3^+}
\def\QR{\mathbb{R}} 
\def\QC{\mathbb{C}} 
\def\a{\alpha}
 
\def\c{\gamma} 
\def\d{\delta} 
\def\e{\epsilon}

\def\cV{{\mc V}} 
\def\cJ{{\mc J}} 
 
\def\ap{{\alpha '}}

\def\bz{{\bar z}}
\def\pl{\partial}
\def\bpl{\bar \partial} 
\def\Ad{{\rm Ad}}

\def\c{\gamma}

\def\cH{{\mc H}}
\def\H{{\mc H}}
\def\cJ{{\mc J}}

\nc{\nn}{\nonumber}
\def\e{\epsilon}

\def\a{\alpha}
 
\def\s{\sigma} 
\def\IP{\relax{\rm I\kern-.18em P}}
\def\tr{{\rm tr\ }}

\def\QC{\mathbb{C}}

\def\QR{\mathbb{R}}

\def\QZ{\mathbb{Z}}

\def\QP{\mathbb{P}}

\def\ppe{\hspace*{0.5mm}}
\def\ew{\hspace*{-1mm}}
\newcommand{\Fus}[6]{F_{{\scriptstyle #1}{\scriptstyle #2}}
  \hspace*{.3mm}\displaystyle{[} \ppe \begin{array}{ll} {\scriptstyle #3 }
  \ppe & {\scriptstyle #4} \ppe \\[-1.5mm] {\scriptstyle #5}\ppe &
  {\scriptstyle #6}\ew \end{array}\displaystyle{]}}

\newcommand{\CG}[6]{\displaystyle{[} \,\ew \begin{array}{lll} 
  {\scriptstyle #1} \ppe
  & {\scriptstyle #2} \ppe & {\scriptstyle #3} \ew \\[-2mm] {\scriptstyle
  #4} \ppe & {\scriptstyle #5}\ppe & {\scriptstyle #6} \ew\end{array}
  \displaystyle{]}}
\newcommand{\SJS}[6]{ \displaystyle{\{ }  \ppe \begin{array}{lll} 
  {\scriptstyle #1} \ppe  & 
  {\scriptstyle #2} \ppe & {\scriptstyle #3}
  \ppe \\[-1.5mm]{\scriptstyle #4}  \ppe & {\scriptstyle #5} \ppe &
 {\scriptstyle #6} \ew \end{array} \displaystyle{\} } }

\def\nn{\nonumber}

\def\cH{{\mc H}}

\def\a{\alpha}

\input epsf
\newcount\figno
\def\fig#1#2#3{
\par\begingroup\parindent=0pt\leftskip=1cm\rightskip=1cm\parindent=0pt
\baselineskip=15pt
\global\advance\figno by 1
\epsfxsize=#3
\centerline{\epsfbox{#2}}
\vskip 12pt
{\bf \small Figure \the\figno:} {\small #1}\par
\endgroup\par
}
\def\figlabel#1{\xdef#1{\the\figno 
\mbox{ }}}
\def\encadremath#1{\vbox{\hrule\hbox{\vrule\kern8pt\vbox{\kern8pt
\hbox{$\displaystyle #1$}\kern8pt}
\kern8pt\vrule}\hrule}}

\def\a{\alpha}
\def\bz{{\bar z}}
\def\cX{{X}}
\def\P1{{\mathbb{P}^1}}

\title{\Large Strings through the Microscope
\footnote{Planary lecture presented at the XIVth International Congress 
on Mathematical Physics, July 28 - August 2, 2003, University of 
Lisbon, Portugal}}

\author{V. Schomerus\\[2mm] 
SPhT CEA/Saclay\\
F-91191 Gif-sur-Yvette\\
FRANCE\\
E-mail: vschomer@spht.saclay.cea.fr\\[3mm]}
\begin{document}
\maketitle 

\begin{abstract}Over the last few years, string theory has changed 
profoundly. Most importantly, novel duality relations have 
emerged which involve gauge theories of brane excitations  on 
one side and various closed string backgrounds on the other. In 
this lecture, we introduce the fundamental ingredients of 
modern string theory and explain how they are modeled through 
2D (boundary) conformal field theory. This so-called `microscopic 
description' of strings and branes is an active research area
with new results ranging from the classification and construction 
of boundary conditions to studies of 2D renormalization group 
flows. We shall provide an overview of such developments before
concluding the lecture with an extensive outlook on some present 
and future research that is motivated by current problems in 
string theory. This includes investigations of non-rational 
and non-unitary conformal field theories.
\end{abstract}
\vspace*{-13cm} {\tt SPhT-T04/055}\hfill\\[13cm] 
\addtolength{\baselineskip}{.6mm}
\pagestyle{plain} 
\section{Introduction}
During the last years several new elements have entered 
the picture of string theory and they have inspired many 
novel ideas in a variety of fields, including high energy physics
and cosmology. The stringy image of our world (see Figure 1) 
contains super-gravity propagating in a 10D background with 
some dimensions being compactified. In addition there are 
branes stretching out along $p+1$-dimensional hyper-surfaces. 
Excitations of these branes give rise to gauge theory 
and matter that can propagate along the brane's world-volume. 
As inspiring as this picture has been for many recent 
developments in physics, it can also be misleading, especially 
when applied to some extreme situations in which the space 
becomes strongly curved or even singular. It is therefore 
crucial to keep in mind that such an image of the world 
only arises in limit of string theories in which the 
involved length scales are long compared to the string 
length and hence that an important task in string theory 
is to develop techniques which allow computing genuinely 
stringy corrections. This goal has been a fruitful challenge 
for more than two decades now and it has lead to many new 
insights within the last years.  
\begin{figure} 
\centering
\scalebox{.4}{\epsfbox{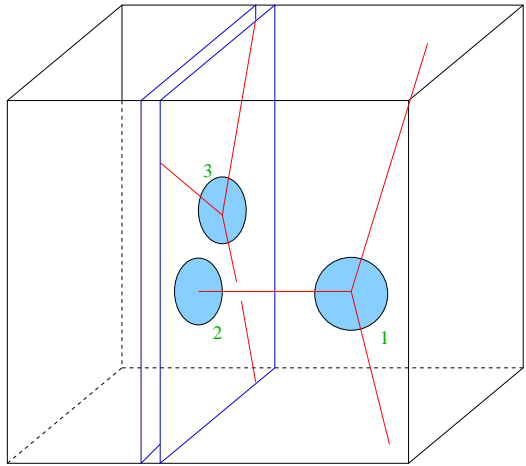}} 
\caption{The picture of modern super-string theory 
contains closed strings which propagate in a 10D background,
$p$-branes stretching along $p+1$-dimensional surfaces and
open strings whose end-points move within the branes' 
world-volume.}  
\end{figure}
\smallskip

Such developments are the main focus of this lecture. 
To begin with, we shall take a much closer look at the various 
elements of Figure 1, enlarging them so that we can see the
strings' finite extension. In the process we shall 
learn how to model the picture in mathematical physics. This
will involve the whole wealth of boundary conformal field 
theory (CFT), a domain that is also known for its beautiful 
applications to critical phenomena (see e.g.\ the lecture of 
Smirnov for some recent applications). As we scan over the 
picture of the world, we shall gradually set up a dictionary 
between its various elements and concepts of 2D boundary 
conformal field theory. 
\smallskip 

After this more introductory part we shall start to discuss 
some of the recent string inspired developments in boundary 
conformal field theory. Our presentation will follow the 
traditional devision into {\em rational vs. non-rational} 
conformal field theory. The former is relevant for the 
study of strings on compact components of the 10-dimensional
world. This subject has seen an enormous boost in recent 
years and by now there exist very powerful techniques 
that are essentially universal, i.e.\ apply to very large 
classes of compact target spaces. Non-rational conformal 
field theory, on the other hand, is relevant for string 
theory in non-compact spaces. As we will discuss, such 
non-compact backgrounds are an essential ingredient in 
string theory duals of (large N) gauge theory. They are 
also vital for studies of time-dependent string 
backgrounds. Applications to these two domains have 
pushed non-rational conformal field theory to the 
forefront of research in string theory.

\section{Strings, branes and boundary conformal field theory}

The main purpose of this section is to zoom in on the elements 
of the picture we sketched above, to understand what they look 
like in full string theory and how they are modeled in 
mathematical physics. 

\subsection{Closed strings and bulk conformal field theory} 

The first element we are going to enlarge is a junction at 
which three closed strings come together (see Figure 2). Our 
aim here is to assign a number to this junction which we can 
interpret as an amplitude for the joining/splitting of closed 
strings. This number depends on the particular background we 
consider and on the states $\phi$ of the three closed strings 
that participate in the process. The latter decorate the three 
external legs of Figure 2 and they are taken from an infinite 
set of possible closed string modes. When we deal with strings 
in flat space, for example, states are characterized by a center 
of mass momentum and an infinite variety of different vibrational 
modes. 
\begin{figure} 
\centering
\scalebox{.4}{\epsfbox{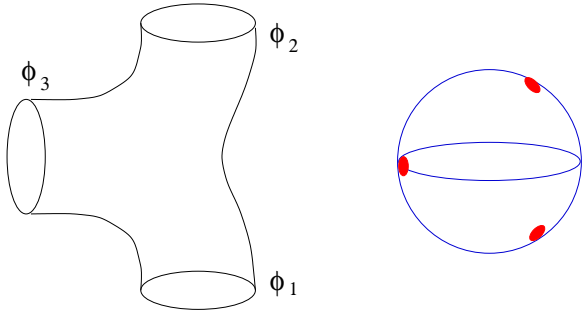}} 
\caption{The interaction 3-vertex of three closed string modes
$\Phi_\nu$ is obtained from a 3-point function of a 2D conformal 
field theory on a closed surface $\QP^1$. The fundamental fields
of the 2D conformal field theory arise through the parametrization 
of the closed string's world-surface.}  
\end{figure}

To assign an amplitude to the vertex, we consider the latter 
as an image of a 3-punctured 2-sphere $\P1$ under the 
parametrization map $(X^\mu(z,\bar z))_{\mu = 0, 1, \dots, 
9}$. Being the parametrization of the closed string's 
world-surface, $X^\mu$ are components of a 10-dimensional 
bosonic field with action 
\be \label{CSA} 
 S_{\rm CS}[X] \ = \ S[X] \ = \ \frac{1}{4 \pi \a'} \, \int \, dz d\bz \, 
  \left(g_{\mu \nu}(X) + B_{\mu \nu}(X)\right) \, \partial X^\mu 
   \bar \partial X^\nu \ + \ \dots 
\ee
where the dots stand for additional terms e.g.\ involving 
Fermions. If our strings propagate in flat space then the 
background metric $g$ and B-field $B$ are constant. This 
implies that the action is quadratic and computations in 
the resulting free field theory can be reduced to Gaussian 
integrals. For more general backgrounds, however, both $g$ 
and $B$ depend on the coordinates of the background and we 
are dealing with (a special class of scale invariant) 2D 
non-linear $\sigma$-models. Our discussion here has brought 
us to the first entry in our dictionary: it is claimed that 
possible {\em closed string backgrounds are associated with 
2D conformal field theories} on closed surfaces. Closed string 
modes $\phi$ correspond to fields $\Phi(z,\bz)$ in the conformal 
field theory. The 3-point functions of such fields 
are determined by conformal symmetry up to some constants 
$$ \langle \Phi_1(z_1,\bz_1) \Phi_2(z_2,\bz_2) \Phi_3(z_3,\bz_3)
   \rangle 
  \ = \ \frac{C(\phi_1,\phi_2,\phi_3)} 
{|z_{12}|^{2\Delta_{12}} |z_{23}|^{2\Delta_{23}}
  |z_{13}|^{2\Delta_{23}}} 
$$
where $z_{ij} = z_i - z_j$ and the exponents $\Delta_{ij}$ are
certain linear combinations of the scaling dimensions $\Delta_i$ 
of the fields $\Phi_i$. Along with the spectrum of these scaling 
dimensions $\Delta_i$, the 3-point couplings $C$ are known to contain 
all the information about the conformal field theory. They obey 
various consistency conditions which are rather difficult to 
solve, but starting from the seminal paper by Belavin et al.\cite{BPZ}, 
many such solutions have been constructed. In string theory, the 3-point 
couplings $C$ provide the amplitude of the 3-point vertex, i.e.\ they 
tell us how likely it is that two closed string modes $\phi_1,
\phi_2$ combine into a single closed string in the mode $\phi_3$. 

\subsection{Branes and boundary conditions in CFT} 
Now that we understand how to model closed strings, let us 
start to look at the next element of Figure 1, namely at  
branes. Originally, the latter entered the image through 
the study of classical super-gravity. In fact, it is known 
for many years that the differential equations of classical  
super-gravity possess static solutions which describe 
charged and massive objects whose mass and charge is 
localized along certain $p+1$-dimensional hyper-surfaces. 
These objects are very much like extremal black holes, 
only that they extend in $p$ spatial directions. Since 
closed string theory is considered as an interesting 
candidate for a consistent short distance completion  
of gravity, we are lead to the obvious problem of
finding a description for branes in string theory. 
\begin{figure} 
\centering
\scalebox{.4}{\epsfbox{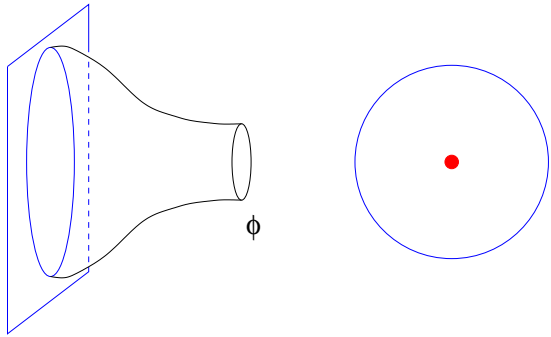}} 
\caption{The interaction of closed strings with a brane is 
related to the bulk 1-point function in a 2D conformal field
theory on a disc. Different branes correspond to different 
boundary conditions imposed along the boundary of the disk.} 
\end{figure}

To answer this question let us now enlarge another element 
of our image (see Figure 3). We claimed above that branes
are charged and massive objects. As such, they can interact 
with various other objects in the bulk, e.g.\ through
exchange of gravitons and higher closed string modes. 
When such a closed string mode hits the brane or is 
emitted from it, we obtain the picture shown in Figure 3. 
The parametrization of the closed string world-surface 
now involves a map from a disc to the 10-dimensional 
background such that the boundary of the disc is 
embedded into the world-volume of the brane. We ensure 
this by imposing Dirichlet boundary conditions for all 
components of $(X^\mu)$ which are associated with 
directions transverse to the brane. Our discussion here 
motivates the following general proposal that was first 
formulated by Polchinski\cite{Pollec}: {\em branes in some 
closed string background correspond to conformally invariant 
boundary conditions} of the associated conformal field 
theory. It is well known that boundary conditions in 
conformal field theory can be characterized by the 
1-point functions of bulk fields $\Phi$. Using once
more conformal invariance and a conformal mapping
from the disc to the upper half-plane, the 1-point 
functions are easily shown to possess the following 
general form 
$$ \langle \Phi(z,\bz)\rangle^{\rm BC} 
    \ = \ \frac{B^{\rm BC}(\phi)}
   {|z-\bz|^{2\Delta_\phi}} \ \ . $$ 
In other words, conformally invariant boundary conditions
are uniquely determined by the 1-point couplings $B(\phi)$. 
The latter provide a measure for how strongly a given closed 
string mode $\phi$ couples to the brane and hence in particular 
encodes information on the mass and charge of the brane.

\subsection{Open strings and boundary fields} 
At this point we are still missing the open strings that we 
would expect to be around as soon as we introduce branes. This
is indeed the case. To argue that open strings are indeed part 
of our present setup, a brief look at lattice spin models with 
boundaries may be helpful. The latter are closely related to the 2D 
continuum field theories we are dealing with.  Moreover, it is 
intuitively obvious that such lattice systems contain a set 
of excitations which can only live along the boundary and which 
depend on the specific boundary condition we impose. In the 
2-dimensional Ising model with free boundary conditions, for 
example, we can measure the boundary magnetization. Once we 
fix the boundary spins, however, measurements of this quantity 
become trivial and hence do not correspond to an observable of
the model. In continuum field theory, boundary excitations are 
described by fields which can only be inserted at points along the 
boundary, i.e.\ to so-called boundary fields. These are 
exactly the objects that we need in order to model open strings. 
As in the case of closed strings, there exists an infinite number 
of {\em open string modes} $\psi$ 
and to these we assign {\em boundary fields} $\Psi(u)$ of the 
corresponding boundary conformal field theory. The spectrum of 
open string modes depends on the brane we consider, 
just as the spectrum of boundary fields depends on the boundary 
condition we impose along the boundary. 
\begin{figure} 
\centering
\scalebox{.4}{\epsfbox{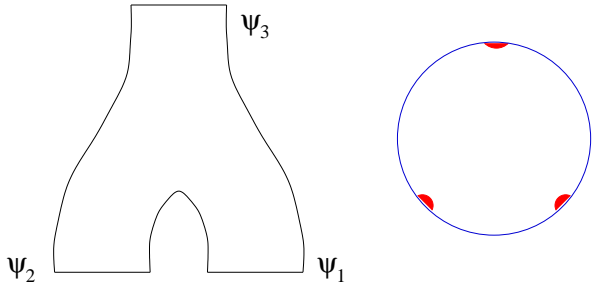}} 
\caption{The interaction 3-vertex of open string modes $\Psi_\nu$ 
is obtained from a 3-point function of boundary fields in a 2D 
conformal field theory on the disc or half-plane.} 
\end{figure}

There is one new set of 
couplings that comes with the vertex of three open string modes 
(see Figure 4). The computation of such a vertex involves a 
disc with three fields inserted along the boundary. Once more, 
this amplitude is determined by conformal invariance up to a 
set of 3-point couplings, 
$$
\langle \Psi_1(u_1) \Psi_2(u_2) \Psi_3(u_3)\rangle 
\ = \ \frac{ O(\psi_1,\psi_2,\psi_3)}
  {|u_{12}|^{\Delta_{12}} |u_{23}|^{\Delta_{23}}
  |u_{13}|^{\Delta_{23}}} 
$$ 
where $u_{ij} = u_i - u_j$ and the $u_i$ are coordinates for 
the boundary of the upper half-plane. Needless to say that  
the 3-point couplings $O$ encode the probability of two 
open string modes $\psi_1,\psi_2$ to combine into a 
single open string in the mode $\psi_3$. 
\smallskip 

This concludes our  
journey through the various elements of Figure 1. Along the way we
have learned to characterize boundary conformal field theories 
through three 
different sets of couplings, the 3-point couplings $C$ of the 
bulk fields, the couplings $B$ appearing in the 1-point function 
of bulk fields and the 3-point couplings $O$ of boundary fields. 
This is a rather abstract way to think about boundary conformal 
field theories, 
especially when compared with the action (\ref{CSA}) for 
closed strings that we started our discussion with. Since it 
is sometimes helpful to think about conformal field theories 
in terms of such 
actions, we would like to mention that in the case of open 
strings, $S[X]$ gets modified to 
\be \label{OSA}
S_{\rm OS}[X] \ = \ S[X] +  \frac{1}{2\pi \ap} \, 
   \int_{\pl \Sigma} du  \,  A_\mu(X)\, \pl_u X^\mu(u)   
\ee
where the first term is the same as in formula (\ref{CSA}) with 
$\Sigma$ being replaced by a surface with boundary. The new boundary 
term means that open strings can couple through the velocity 
$\pl_u X^\mu$ of their endpoints to a new background field, 
namely to a vector field $A_\mu(X)$. We interpret $A$ as a 
gauge field on the brane and think of the string endpoints 
as carrying charges. The action (\ref{OSA}) is to be supplemented 
by Dirichlet boundary conditions on $X$ for all directions
transverse to the brane.

\subsection{Instabilities, dynamics and RG-flows} 

In the last three subsections, our dictionary has received 
quite a number of entries, but there are a few more that we 
would like to add in passing. To motivate these additional entries,
we note that many of the existing string backgrounds are 
actually unstable. This applies e.g.\ to the 26-dimensional 
bosonic string theory and it is a common phenomenon in the 
context of branes. In fact, configurations of several stable 
branes tend to be unstable, the most famous example being 
the pair of a brane and its anti-brane. 
\smallskip 

In general relativity we can easily detect instabilities of
some given static solution to the classical field equations. 
All we need to study is the spectrum of fluctuations around
the solution. Modes that are exponentially enhanced are 
associated with instabilities. Let us now see
how we can find unstable modes in string theory. To this end
we consider some static background. Its conformal field theory 
description involves a product of a time-like free field $X^0$ 
and a unitary conformal field theory whose target is the 
9-dimensional spatial slice of the static background. This
theory possesses an exponentially growing mode if we can 
add the following conformally invariant terms to the 
action 
\begin{equation} \label{tachact}  
 \delta S \ \sim \ \int_\Sigma  dzd\bz\   
     e^{\epsilon_\Phi X^0(z,\bz)}\, \Phi(z,\bz) + 
  \int_{\pl \Sigma}  du \ e^{\epsilon_\Psi X^0(u)}\, 
   \Psi(u)      
\end{equation} 
where $\Phi(z,\bz)$ and $\Psi(u)$ are bulk and boundary 
fields in the conformal field theory for the spatial slice 
and the coefficients $\epsilon_\Phi$ and $\epsilon_\Psi$ 
must be real in order to 
have an exponential growth with time. But such exponential 
fields of a time-like free boson possess a positive scaling 
dimension so that the fields $\Phi$ and $\Psi$ must have 
scaling dimensions $\Delta_\Phi \leq 2$ and $\Delta_\Psi 
\leq 1$ if we want $\delta S$ to be scale invariant. In 
other words, instabilities of a closed string background 
and branes therein correspond to relevant bulk and boundary 
fields, respectively.%
\smallskip 

If we wanted to control the full decay process generated by an 
instability, 
we would have to solve the theory with interaction $\delta 
S$. At present, no example of such a theory has been constructed 
within 10D string theory. Nevertheless, some insights into aspects of
decay processes have been obtained using a proposed relation with 
renormalization group (RG) flows. This requires, however, that we
reduce our ambitions and only try to identify the final state of
the decay process rather than the whole dynamical evolution. Let 
us consider an initial state which is encoded in a CFT$_1$ for the 
9-dimensional spatial slice of the static background. Next we 
choose some relevant bulk or boundary field $\Phi$ or $\Psi$
to trigger the decay. After all radiation has has escaped, we
expect our system to settle down in a static final state that 
is again described by a product of a time-like free boson and 
a CFT$_2$ for the spatial slice of the final stable state. 
Hence, the dynamics has lead us from some CFT$_1$ to CFT$_2$ 
with the help of relevant fields $\Phi$ or $\Psi$. It is 
obviously tempting to think that CFT$_2$ is the IR fixed 
point of the RG flow from CFT$_1$ on the RG trajectory that is
generated by adding $\Phi$ or $\Psi$. Supporting evidence for
this proposal is strong in the case of boundary instabilities, 
but the proposal seems much harder to justify when we deal 
unstable bulk theories. A more thorough discussion of these 
issues and many further references can be found in the 
literature\cite{Gutperle:2002ki}.

\subsection{Summary: dictionary between ST and BCFT}  
\def\cX{{\mc X}}
\def\cB{{\mc B}}

Let us pause for a moment and review all the entries in 
our dictionary between string theory and 2D boundary 
conformal field theory: \\[.5cm] \hspace*{1cm}
\begin{tabular}{|c|c|} \hline & \\[-3mm]   \hspace*{.5cm}
Closed string background $\cX$  \hspace*{.5cm}
            &  2D CFT $\cC_{\cX}$ on closed surface $\Sigma$ 
\\[1mm]  \hline & \\[-3mm]  
Closed string mode $\phi$       
            &    Bulk field $\Phi(z,\bz), z \in \Sigma,$ in 
                   $\cC_\cX$  
\\[1mm] \hline &\\[-3mm]  
Closed string vertex $C$    
            &    3-point function of bulk fields 
\\[1mm] \hline &\\[-3mm]  
Brane $\cB$ in background  
            &    Boundary condition (BC) for $\cC_\cX$ \\[2mm]
            &    $\Rightarrow$ BCFT $\cC_{\cX}^\cB$ on open 
                 surface $\Sigma$
\\[1mm] \hline &\\[-3mm]  
Open string mode $\psi$ 
            &    Boundary field $\Psi(u), u \in \pl \Sigma,$ in 
                 $\cC_\cX^\cB$   
\\[1mm] \hline &\\[-3mm]  
Open string vertex $O$
            &   3-point function of boundary fields 
\\[1mm] \hline &\\[-3mm]  
Instability of a background             
            &   Relevant bulk- or boundary-field
\\[1mm] \hline &\\[-3mm]  
{\em 
Initial/final state of decay} 
            &    {\em UV/IR fixed point of an RG-flow} \\[1mm] 
\hline  \end{tabular}  \\[5mm]
\noindent 
Recall that the last line has the status of a conjecture. 
To test our dictionary one may form sentences in string 
theory (or gravity) and check that they translate into 
meaningful sentences of (boundary) conformal field 
theory.

\section{Rational BCFT and strings in compact backgrounds}    

We are now prepared to begin reviewing results on the explicit
construction of boundary conformal field theories. Systematic 
rational conformal field theory model building usually starts 
with theories whose target space is a compact group manifold 
$G$. It then proceeds to cosets and orbifolds. Among them one 
finds all known models with interesting applications to 
statistical physics and string theory. Here we shall explain 
the most central results of the field using one special 
example, namely the group $G=$SU(2)$\cong S^3$. A few comments 
on various generalizations and extensions are collected 
at the end.

\subsection{Strings on the 3-sphere} 

Before we look at strings moving on a 3-sphere, let us recall that, 
to leading order in $\ap$, $\s$-models of the form (\ref{CSA}) are
conformal invariant if the background fields satisfy
$$  \ap \cR_{\mu\nu}(g) - \frac{\ap}{4} H_{\mu\lambda\omega} 
     H_\nu ^{\ \lambda \omega} \ = \ O(\ap^2)\ \ .      $$ 
Here, $\cR$ is the curvature of the background metric $g$ and $H = dB$. 
Hence, if strings move in a curved background, then a non-vanishing 
magnetic field $B$ is unavoidable. 
\smallskip 

With this in mind let us consider strings on a 3-sphere $S^3 
\cong$ SU(2). Their world-surfaces are parametrized by a 
group valued map $h: \Sigma \rightarrow$ SU(2) and these 
parametrization fields appear in an action of the form 
\begin{eqnarray}
S[h] \ = \ \frac{\ik}{4\pi} \int dz d\bz \ 
    \tr(h^{-1} \partial h)  (h^{-1} \bar \partial h)  
      + \frac{\ik}{12\pi} \int d^{-1} 
                (h^{-1} dh)^{\wedge_3} \ \ .    
\end{eqnarray}   
The second term is known as the Wess-Zumino-Witten (WZW) term. 
Locally, it may be rewritten in the form of the second term in 
eq.\ (\ref{CSA}). We also note that the parameter $\ik$ is a 
measure for the size of the 3-sphere and that, in the quantum 
theory, $\ik$ must be integer.  
\smallskip 

The WZW model possesses a large symmetry, given by  two commuting 
actions of the affine Lie algebra  ${\widehat{{\rm su}}(2)}_\ik$. 
Each of these two algebras is generated by the Laurent modes 
$J^\mu_n, n \in \QZ,$ of an su(2)-valued conserved current $J 
= J^\mu t_\mu$ with relations
\be \label{KMcomm}
[ \, J^\mu_n \, , \, J^\nu_m \, ] \ = \ 
 i {f^{\mu\nu}}_\rho\, J^\rho_{n+m} + {\ik} \, n 
  \, \d^{\mu,\nu} \,   \delta_{n,-m} \ \ . 
\ee 
The two affine Lie algebras act on the space of fields in the theory,  
extending the actions that are induced by the usual left and right 
translation of the group on itself.  
\smallskip 

It turns out that a wide class of boundary conformal field theories 
can be written down in terms of data from the representation theory 
of their infinite dimensional symmetries. The most important such 
data are the set $\cJ$ of unitary representations, the so-called 
modular $S$-matrix, the Clebsch-Gordan multiplicities $N$ of the 
fusion product and an infinite dimensional generalization of the 
6J-symbols which is known as the Fusing matrix $F$. For the SU(2) 
affine Lie algebra, explicit formulae may be spelled out. In this 
case, the requirement of unitarity leaves us 
with just a finite number of representations. We label them through 
$j \in \cJ = \{0,1/2, \dots, \ik/2\}$. Their conformal weights are 
given by $\Delta_j = j(j+1)/(\ik+2)$ and for the modular S-matrix one 
finds
\be \label{Smatrixsu}
S_{ij} \ = \ \sqrt{\frac{2}{\ik+2}} \ \sin \frac{\pi(2i+1)(2j+1)}{\ik+2} 
\ \ . 
\ee
With the help of the Verlinde formula it is not difficult to compute 
the following fusion rules $N$ from the modular S-matrix, 
\be {N_{ij}}^k \ = \ \left\{ \begin{array}{ll} 
   1 \ \ \ \ & \mbox{ for } \ k = |i-j|, \dots, \min (i+j,\ik-i-j)
   \\[2mm] 0 & \mbox{ otherwise } 
   \end{array} \right. \ \ .   
\label{wzwfus} \ee
They are similar to the Clebsch-Gordan multiplicities of the Lie 
algebra su(2), apart from the truncation which appears whenever 
$i+j > \ik/2$.   
\smallskip 

Formulae for the fusing matrix also exist in the literature. Since 
they are a bit more involved, we shall not present them here. Let 
us only mention one property concerning their limiting behavior as 
we send $\ik \rightarrow \infty$, 
\be \label{fusprop} 
\lim_{\ik \rightarrow \infty} \Fus{J}{k}{i}{j}{I}{K}
\ = \ \SJS{i}{j}{k}{I}{J}{K}       \ \ .     
\ee
This concludes our list of representation theoretic data for 
the affine Lie algebra. We shall see these quantities again in 
a moment when we write down formulae for the couplings $B$ of 
closed strings to branes on $S^3$, for their open string spectra 
and the 3-point vertices $O$ of open string modes. 

\subsection{Branes on group manifolds} 

Many constructions of conformal invariant boundary theories are
ultimately based on fundamental observations made by J.\ Cardy%
\cite{Card89}. When applied to the case at hand, they provide us 
with a finite set of $\ik+1$ boundary conformal field theories 
which we label by $J = 
0,1/2, \dots,\ik/2$, just as we enumerate the unitary representations 
of the corresponding affine Lie algebra. Recall that these boundary 
theories can be uniquely characterized through the couplings $B$ 
that appear in the 1-point functions of the bulk fields. 
According to Cardy's solution, these 1-point couplings are simply 
given by the matrix elements of the modular S-matrix, 
\be \langle \Phi_{j}^{ab}(z,\bz) \rangle_J \ = \ 
    \frac{S_{Jj}}{\sqrt{S_0j}} 
    \ \frac{\delta^{a,b}}{|z-\bz|^{2\Delta_j}}
\label{wzw1pt} \ee
where $\Delta_j = j(j+1)/(\ik+2)$. The superscripts $a,b = -j, \dots, 
j,$ placed at the symbol $\phi$ label different components within 
a tensor multiplet $\phi_{j}$. One should think of these labels 
as the quantum numbers for ground states of the closed string. They 
are the same labels that appear on the matrix elements $D^j_{ab}(g)$ 
of group representations, except that the label $j$ is cut off at 
$j \leq \ik/2$.   
\smallskip
 
Geometrically, these boundary theories were found\cite{AleSch5} to 
describe branes that are localized along a discrete set of conjugacy 
classes of SU(2), i.e. along 2-spheres around the group unit $e \in$ 
SU(2). They come equipped with a magnetic field of the form 
\be
B \ \sim \ \tr h^{-1} dh \, \wedge \, \frac{\Ad_h + 1}{\Ad_h -1} 
   \, h^{-1} dh \ \ , 
\ee
where $\Ad$ denotes the adjoint action of the group on its Lie
algebra. For higher groups $G$, such a field $B$ gives a non-trivial
potential for the pull-back of the WZW 3-form to the branes' 
world-volumes. Such 2-form potentials on conjugacy classes of a group 
$G$ were considered in the mathematical literature. There they appear 
in connection with deformations of the theory of co-adjoint orbits. 
\begin{figure} 
\centering
\scalebox{.4}{\epsfbox{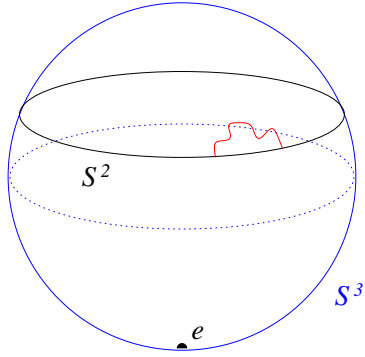}} 
\caption{Maximally symmetric branes on a 3-sphere are localized 
along conjugacy classes of SU(2), i.e.\ they are either point-like
or they wrap a 2-sphere in $S^3$. Such spherical branes exist only
for a discrete set of radii.} 
\end{figure}
    
Our geometrical interpretation of the boundary theories with 1-point 
functions (\ref{wzw1pt}) may be justified in several different ways. 
The argument we want to give here is based on the 
observation\cite{FFFS1}     
\be \label{Lkc} 
     \langle \Phi^{ab}_{j}(z,\bz) \rangle_{J(\ik)} 
   \ \stackrel{\ik \rightarrow \infty}{\sim}  \ 
   \int_{\rm SU(2)} d\mu(g) \, \delta(\vartheta(g)-
     \vartheta_0)\, \phi^{ab}_j(g) \ \ ,  
\ee 
where $d\mu(g)$ denotes the Haar measure on SU(2), $\phi_j^{ab}(g) 
\sim  D^j_{ab}(g)$ are the wave functions of the lightest closed 
string modes and $\vartheta$ is the azimuthal angle on the 3-sphere.    
In taking the limit, we allowed the boundary label $J$ to depend 
on the level $\ik$ and we defined $\vartheta_0 := 2 \pi \lim 
J(\ik)/\ik$. Since $0\leq J \leq \ik/2$, the angle $\vartheta_0$ 
lies in the interval $[0,\pi]$. The appearance of the 
$\delta$-function on the right hand side shows that closed 
string modes indeed detect a spherical object which is 
localized at the azimuthal angle $\vartheta = \vartheta_0$.

\subsection{Open strings on group manifolds} 
\def\cH{{\mc H}}

Let us now turn to the open strings which can propagate along 
the $\ik+1$ different spherical branes on $S^3$. According to 
our dictionary, this means that we want to obtain the set of 
boundary fields and the 3-point couplings $O$ for each of the
above boundary theories. Following general arguments, it is 
possible to show that the space of boundary fields carries 
the action of a single affine Lie algebra rather than two 
commuting actions as in the case of bulk fields. This reflects 
a similar reduction of the geometric symmetries: whereas there 
exist two commuting sets of (left and right) translations on SU(2), 
only a special combination, namely the adjoint action, leaves the 
conjugacy classes invariant. Under the action of the affine 
Lie algebra ${\widehat{{\rm su}}(2)}_\ik$, the space $\cH_J$ 
of boundary fields decomposes as 
\be \label{partdec}
 \cH_{J} \ = \ \cH_{JJ} \ = \ {\bigoplus}_j \ {N_{JJ}}^j \ \cV_j 
\ee
where $\cV_j$, $j = 0,1/2, \dots,\ik/2$, denote irreducible 
unitary representations of the affine Lie algebra ${\widehat{{\rm su}}
(2)}_\ik$, and where ${N_{JJ}}^j$ are the associated fusion rules (see 
eq.\ (\ref{wzwfus})). Note that only integer spins $j$ appear on the 
right hand side of eq.\ (\ref{partdec}) and that in the limit $\ik 
\rightarrow \infty$, the summation on the right hand side is truncated 
at $j_{max} = 2J$. This means that the decomposition of $\cH_{J}$ is 
as close as it can be to the decomposition of $\Mat(2J+1)$ into su(2) 
multiplets. In fact, there is a correspondence between the spin $j$ 
multiplets of matrices $\rmY^j_a \in \Mat(2J+1)$ and ground states 
in $\cV_j\subset \cH_J$. It is also worth pointing out the similarity
between the labeling of open string ground states and spherical 
harmonics $Y_a^j(\varphi,\theta)$. Only the cut-off at $j_{max} = 2J$ 
on the spin $j$ does not appear for functions on a 2-sphere.   
\smallskip

Boundary fields $\psi^a_j$ associated with ground states of the open 
string are labeled by the representation $j = 0,1, \dots, j_{max}$ and
$a = -j, \dots,j$. Their 3-point vertices were found\cite{AlReSc2} 
using important previous work by Runkel\cite{Runk1} on minimal 
models,  
\be \label{boundOPE}
  \langle \Psi^a_i(u_1)\, \Psi_j^b(u_2) \, \Psi_k^c(u_3)
   \rangle_J  \ = \   {|u_{12}|^{-\Delta_{12}} |u_{23}|^{-\Delta_{23}}
  |u_{13}|^{-\Delta_{23}}}  
      \,  \CG{i}{j}{k}{a}{b}{c}\, \Fus{J}{k}{j}{i}{J}{J}    
\ee
where the symbol in square brackets stands for the 
Clebsch-Gordan coefficients of su(2). The latter guarantee 
that both sides of the equation transform in the same way 
under the action of the zero mode algebra su(2). The non-trivial 
part of equation (\ref{boundOPE}) therefore concerns the relation 
of 3-point vertices for open strings with entries of the Fusing 
matrix. 
\smallskip 

We conclude this subsection with a brief remark on an interesting 
link to non-commutative geometry. Let us observe that the 3-point 
couplings simplify significantly if we send $\ik$ to infinity, 
\be \label{boundOPEtop}
    \langle \Psi^a_i(u_1)\, \Psi_j^b(u_2)
     \, \Psi_k^c(u_3)\rangle_J^{\ik \rightarrow \infty} 
    \, = \, 
    \SJS{j}{i}{k}{J}{J}{J} \, \CG{i}{j}{k}{a}{b}{c} \ \ . 
\ee
Here we have used the property (\ref{fusprop}) of the fusing 
matrix. It is straightforward to check that the numbers appearing 
on the right hand side of this equation arise naturally from the
multiplication of ordinary $(2J+1)\times(2J+1)$ matrices. In fact, 
the same numbers appear when we rewrite the product $\rmY^i_a 
\cdot \rmY^j_b$ as a linear combination of the su(2) multiplets 
$\rmY^k_c \in \Mat(2J+1)$. In this sense, the 3-point vertices 
for open strings on branes in SU(2) provide an infinite dimensional 
deformation of matrix multiplication. The emergence of non-commutative 
matrix algebras in the context of open string theory is not surprising. 
As we have stressed before, the end-points of open strings behave like
charged particles which can couple to the vector potential $A$ of the 
magnetic field on the brane. Hence, the non-commutativity we 
encounter in the context of open strings is directly related to 
a similar phenomenon for e.g.\ electrons in a strong magnetic 
field. Relations between branes, open string and non-commutative 
geometry were initially discovered for branes in flat 
space\cite{DouHul,ChuHo1,Scho} and they have been studied 
extensively for several years.\cite{SeiWit99,Douglas:2001ba}

\subsection{Brane dynamics on group manifolds} 

Before we list a few generalizations of the above results,
we would like to briefly comment on some dynamical processes
involving branes in $S^3$. Here we shall use the conjectured 
relation with RG-flows in the 2D boundary conformal field 
theory. It turns out that the study 
of boundary renormalization groups flows in models with an 
SU(2) current algebra is a classical problem of mathematical 
physics that was first addressed in the context of the 
Kondo-model. 
\smallskip 

The Kondo-model is designed to understand the effect of magnetic 
impurities on the low-temperature conductance properties of a 3D 
conductor. The latter can have electrons in a number $\ik$ of 
conduction bands. If the impurities are far apart, their effect 
may be understood within an s-wave approximation of scattering 
events between a conduction electron and the impurity. This allows 
to formulate the whole problem on a 2-dimensional world-sheet for 
which the coordinates $(u,v)$ are associated with the time and
the radial distance from the impurity, respectively. One can 
build several currents out of the basic fermionic fields. Among 
them is a {\em spin current} $\vec{J}(u,v)$. Its Laurent modes 
satisfy the relations (\ref{KMcomm}) of a $\widehat{\rm su}(2)_\ik$ 
current algebra. This spin current is the one that couples to the 
magnetic impurity of spin $J_M$ which is sitting at the boundary 
$v=0$,    
\begin{equation} \label{Kondo} 
 S_{\rm Kondo} \ \sim \ \lambda \int_{-\infty}^\infty du\, 
\Lambda_\mu J^\mu(u,0)  \ \ . 
\end{equation}
Here, the matrices $\Lambda_\mu, \mu=1,2,3$ form a  
$(2J_M+1)$-dimensional irreducible representation of su(2)
and the parameter $\lambda$ controls the strength of the coupling.  
Note that the term (\ref{Kondo}) is identical to the coupling of 
open string ends with velocity $J^\mu(u,0)$ to a background gauge 
field $A_\mu = \Lambda_\mu$ (see formula (\ref{OSA})). Hence, 
$\Lambda_\mu$ may be interpreted as a constant non-abelian 
gauge field on one of our branes. For $M > 1$, the interaction 
is marginally relevant so that, according to the proposal 
formulated in subsection 2.4, switching on such non-abelian 
gauge fields represents an instability. Its effect on the  
brane can be understood by searching for an RG fixed-point 
along the RG trajectory generated by the term (\ref{Kondo}).          
\smallskip 

Fortunately, a lot of techniques have been developed to deal with 
perturbations of the form (\ref{Kondo}). From the old analysis it 
is known that  
non-trivial fixed points are reached at a finite value $\lambda = 
\lambda^*$ of the renormalized coupling constant $\lambda$ if 
$ 2 J_M \leq \ik$ (exact- or over-screening resp.). These fixed 
points have been identified through several different approaches
and a simple rule summarizing the results of such investigations 
was formulated by Affleck and Ludwig\cite{Affleck:iv}. The latter 
can be applied directly to our branes on SU(2)\cite{Fredenhagen:2000ei} 
and it shows that e.g.\ point-like branes carrying a constant U(M) 
gauge field decay into an extended spherical brane with label 
$J_M = (M-1)/2$.     
\begin{figure} 
\centering
\scalebox{.4}{\epsfbox{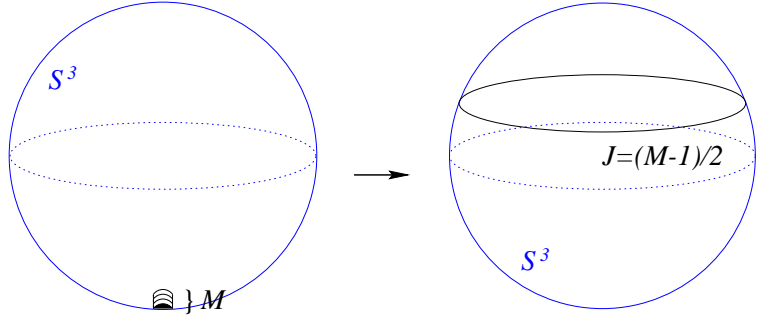}} 
\caption{Studies of RG flows in models with a SU(2) current symmetry 
show that point-like branes carrying a constant U(M) gauge field decay 
into an extended spherical brane with label $J_M = (M-1)/2$.} 
\end{figure}
  
For large values of $\ik$, the result we have just formulated 
admits an easier derivation using some of the geometric 
structures we have outlined above. It is widely known that 
massless open string modes give rise to a gauge theory on the 
brane to which they are attached. But due to the presence of 
the $B$-field on our spherical branes, the open strings
detect a non-commutative (matrix) geometry (cf. last subsection). 
Hence, we are tempted to conclude that a special non-commutative 
gauge theory should be associated with branes on the 3-sphere. 
This is indeed the case and the precise form of the relevant 
gauge theory has been determined following the usual rules of 
string theory\cite{AlReSc3}, at least to leading order in $\a'$. 
Once the action of this gauge theory is found, one can search 
for instabilities and classical ground states. The outcome of 
such an analysis confirms very nicely the formation of spherical 
branes from point-like branes that we have described in our 
discussion of RG-fixed points above. This agreement is not 
accidental. In fact, when $\ik$ is large, the action of the 
non-commutative field theory is a functional on the space 
of boundary couplings whose stationary points approximate 
zeros of the $\beta$-function. The analysis of RG-fixed points 
through non-commutative field equations has significant 
technical advantages over more traditional approaches. These  
are related to a very efficient book-keeping of the possible 
boundary couplings through non-commutative variables.

\subsection{Overview: various generalizations and extensions} 
 
The material presented in this section has been generalized 
in many different directions and we would like to list a few 
of these developments before we leave the rational boundary 
conformal field theories.\footnote{The references provided 
in the following 
paragraph are highly incomplete. A more extensive list can 
be found e.g.\ in the lecture notes\cite{Schomerus:2002dc}.} 
Not surprisingly, results similar to the 
ones we have reviewed here are also available for all other 
compact simple simply connected Lie groups. For many of the 
higher groups, moreover, there exist new families of maximally 
symmetric branes that are associated with outer automorphisms 
of $G$ and hence have no analogue in the case of SU(2). Their 
boundary couplings $B$ and open string spectra were first 
studied by Felder et al.\cite{FFFS1} (see also \cite{PetZub02,%
GabGan}). Results on the open string couplings and dynamics 
have also been obtained\cite{AFQS}. In addition to the maximally 
symmetric branes, i.e.\ those that admit an action of the whole 
group $G$, one can impose boundary conditions which break part 
of this symmetry. Investigations of such branes were initiated 
in a paper by Moore et al.\cite{MaMoSe1} and a rather systematic 
construction has been developed more recently\cite{QueSch1}. It is 
worth mentioning that symmetry breaking branes on group manifolds 
possess interesting applications to defect lines in 2D conformal
field theory. With 
the theory of strings and branes on group manifolds being under 
such good control, one can start to descend to orbifolds and 
cosets thereof. Studies of branes and open strings in curved 
orbifold backgrounds possess a long history\cite{Pradisi:1988xd,%
Bianchi:1990yu,AffOsh2,Fuchs:1999zi,BiFuSc}. The theory of branes in 
coset models has also been treated by many authors (see e.g.\ 
\cite{MaMoSe1,Gawe2,EliSar,FreSch3} for some early 
contributions and references).%

\section{Recent progress for strings in non-compact spaces}

The last part of this lecture is devoted to some developments 
in the area of non-rational boundary conformal field theory. 
As indicated in the introduction, these are highly relevant 
for the study of dualities between string and gauge theory 
and for time-dependent processes in string theory. Here we 
shall explain these motivations in some detail and then 
focus mainly on one particular model, namely on the 
Liouville field theory.%

\subsection{String/gauge theory dualities} 

According to an old observation by 't Hooft, structures 
found in closed string amplitudes are very reminiscent 
of features in the perturbative expansion of large $N$ 
gauge theories. This suggests that it might be possible 
to compute gauge theory amplitudes from string theory. 
Though the idea has been around for a long time, only 
a single concrete example was known until 1997: The
duality between the quantum mechanics of hermitian 
matrices and a toy model of string theory with a 
2-dimensional target space (see e.g.\ the lectures 
of Klebanov\cite{Klebanov:1991qa} and references 
therein).%
\smallskip 

After branes had entered the stage of string theory, the 
situation changed drastically. We may grasp the 
fundamental role of branes for such developments through 
the following short analysis of Figure 7. The image shows 
a simple string diagram that admits two rather distinct
interpretations. We can either think of a process in which 
a closed string mode is emitted from one brane and propagates 
a distance $\Delta x$ through the background before being 
absorbed by a second brane. Alternatively, the process can 
be seen as a pair creation and subsequent annihilation of 
open strings which end on the two involved branes. Though 
none of these interpretations is distinguished a priori, 
actual computations may favor one of them, depending 
on the separation $\Delta x$ between the branes. If they  
are very far apart only the massless closed string modes 
contribute significantly to the amplitude and hence one 
can safely perform the string theory computation in its 
super-gravity approximation. For short distances $\Delta x$ 
between the branes, however, many massive closed string 
modes start to enter the computation and we better switch 
to a description in terms of light, i.e.\ mildly stretched, 
open strings. This reinterpretation leaves us with a one-loop 
computation in the gauge theory of massless open string modes. 
The reasoning we have just gone through hints toward an intimate 
relation between closed string models and gauge theory. This 
relation has a few interesting features. Observe, for example, 
that it does not preserve the underlying space-time dimensions since
closed strings propagate in the 10D background while gauge bosons 
cannot leave the branes' $p+1$-dimensional world-volume. In 
addition, the relation also mixes different loop orders as 
can be inferred from Figure 7. Here, we found a closed string 
tree level amplitude that contains information about a gauge 
theory one-loop diagram.%
\begin{figure} 
\centering
\scalebox{.5}{\epsfbox{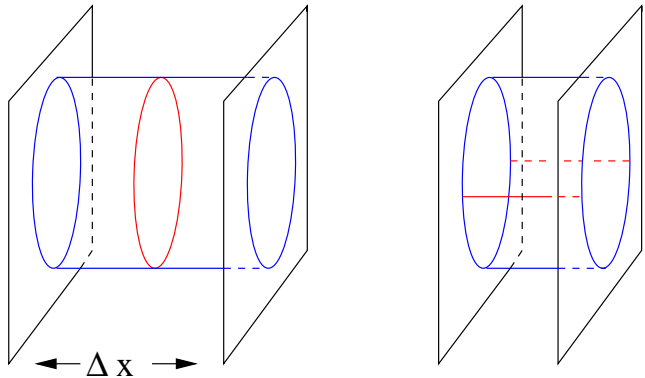}} 
\caption{Depending on the distance $\Delta x$ between branes
the shown string amplitude can be approximated by either a 
tree level computation in super-gravity or a one-loop gauge
theory computation.}  
\end{figure}

With all this in mind it seems no longer surprising that 
branes provide us with many new and concrete dualities between 
gauge theories and models of closed strings. The most famous 
example certainly is Maldacena's duality\cite{Maldacena:1997re} 
between ${\mc N}=4$ Super-Yang-Mills theory and string theory 
on $AdS_5 \times S^5$. It involves an SU(N) gauge theory at 
large N with 't Hooft coupling $\lambda = g_{\rm YM} N^2$ on 
one side and an $AdS_5$ space with curvature radius $R^2/\ap
\sim \sqrt\lambda$ on the other. Hence, if we want to study gauge 
theory at finite 't Hooft coupling, the curvature of $AdS_5$ 
cannot be neglected so that string effects become important. 
Not only does this bring us back to the main theme of the  
lecture, it also hides novel challenges arising from the 
fact that the relevant closed strings propagate on a 
non-compact curved background. Similar observations can
be made for all the other known examples of such dualities 
between gauge  and string theory\cite{AGMOO}. Therefore, 
extending the methods and results reviewed in the previous 
section to non-compact target spaces becomes an important 
new task for conformal field theory.%
\smallskip 

Here we shall restrict our attention to the example of  
Liouville theory. Since it appears as constituent of 2D 
string theory, Liouville theory enters one side of the 
aforementioned duality with matrix quantum mechanics. 
More recent developments show that even this duality, 
though it pre-dates the discovery of branes in string 
theory by several years, ultimately arises is the same 
way as all the modern AdS/CFT-type dualities.%

\subsection{Liouville theory and its applications} 

Liouville theory may be considered as the minimal model 
of non-rational conformal field theory. It describes the 
motion of strings in a single direction with an 
exponential potential
\be \label{actLiouv} 
   S_L[X] \ \sim \ \int_\Sigma dz d\bz \, 
     \left( \pl X \bpl X + \mu e^{2bX} \right)  
\ \ . 
\ee
Here, $X$ is a free bosonic field with background charge
(linear dilaton). Note that a change of the parameter 
$\mu$ can be re-absorbed in a constant shift of the field 
$X$. Hence, the dependence of Liouville theory on $\mu$ is 
rather trivial. On the other hand, the correlation functions
of the model display a very non-trivial dependence on the 
second parameter $b$. If we send $b$ to zero while rescaling 
$x = bX$ and $\lambda = b^2 \mu $, we recover a model of 
particles in the potential $V(x) = \lambda \exp 2x$. The 
full field theoretic model contains both perturbative and 
non-perturbative corrections to this limit. The latter turn 
out to render the quantum theory self-dual with respect to 
the replacement $b \rightarrow b^{-1}$.    
\smallskip 

After these remarks on the model itself, let us briefly 
comment on two interesting applications. It is well 
known that a flat space with trivial 
background fields must be 26-dimensional for bosonic 
strings to propagate in it. This restriction to 26 
dimensions, however, may be circumvented through the 
presence of non-trivial background fields. An example
realizing such a scenario is the famous 2D string 
theory. It is constructed from a product of a 1-dimensional 
free bosonic field with Liouville theory at $b =1$. Obviously, 
such a closed string theory is of little practical interest, 
but is provides a valuable toy model for bosonic string 
theory. Compared to its 26-dimensional relative, the 
2-dimensional model has the advantage of being 
perturbatively stable. 
\smallskip 

The second application is much more recent and also less well tested. 
It has been proposed\cite{Gutperle:2003xf} that a time-like version 
of Liouville theory at $b=i$ describes the homogeneous condensation 
of a closed string tachyon. A short look at the classical action 
(\ref{tachact}) makes this proposal seem rather plausible. Recall 
that actions of this form describe a tachyon decay with a closed/open 
string tachyon profile $\Phi/\Psi$. If we consider a 
closed string tachyon which decays homogeneously, then we must 
choose $\Phi = \mu =${\it const}. In addition, the parameter 
$\epsilon_\phi$ is now forced to be $\e_\mu = 2$ so that the 
interaction term (\ref{tachact}) becomes scale invariant. After 
a Wick-rotation of the field $X^0 = i X$, the interaction term 
looks formally like the interaction term in Liouville theory, 
only that the we have $b = i$, just as it was claimed above. 
Though on first sight this may all seem rather convincing, 
there are many subtleties hidden in the relation between 
Liouville theory and tachyon condensation. We will be able 
to display them more clearly once we have presented the 
exact solution of Liouville theory.

\subsection{The solution of Liouville field theory} 
\def\ta{\tilde \alpha}

The solution of Liouville theory on closed and open surfaces 
has been a major success of the last ten years. We shall 
briefly spell out some of the most central formulae before we 
return to the applications. 
\smallskip 

Let us begin with a few simple remarks on the spectrum of 
the model. It is rather obvious that ground states $\phi_P$ 
of closed strings in the exponential potential can be labeled 
by a real number $P \geq 0$ which corresponds to the momentum 
of an incoming wave in the region $X \rightarrow -\infty$ where 
the potential vanishes. The states $\phi_P$ are stringy analogues 
of the particle wave functions 
$$ \varphi_P(x) \ = \  (\lambda/4)^{-i\omega}
   \, \Gamma^{-1} (-2i\omega) \, 
   K_{2i\omega} (\sqrt{\lambda} e^x) $$ 
for a Schroedinger problem in an exponential potential. Here, 
$K$ denotes the modified Bessel function of second kind.  
Finding the 3-vertex of the closed string modes $\phi_P$
turned out to be a very difficult problem that was only 
solved in 1994. The answer is usually expressed in terms
of $\a_j = Q/2 + i P_j$ where $Q = b + b^{-1}$, 
\be \label{CSL}   
 C(\a_1,\a_2,\a_3) \ = \ \left[\pi \mu \gamma(b^2) (b^2)^{1-b^2}
\right]^{(2\ta - Q)/b} \  \frac{\Upsilon'(0)}{\Upsilon(2\ta - Q)} 
\ \prod_{j=1}^3 \,  \frac{\Upsilon(2\a_j)}{\Upsilon(2\ta_j)}\ \ .
\ee
Here, we have also introduced the notation $\ta = (\a_1 + \a_2 + 
\a_3)/2$ and $\ta_j = \a- \a_j$. In addition, the special function 
$\Upsilon$ is obtained from Barnes' double Gamma function by 
\beq\label{Y}  
\Upsilon (\a) \ :=\ \Gamma^{-1}_2(\a|b,b^{-1}) 
\, \Gamma^{-1}_2(Q-\a|b,b^{-1}) \ \ . 
\eeq 
The solution (\ref{CSL}) was first proposed by H.\ Dorn and 
H.J.\ Otto \cite{Dorn:1994xn} and by A.\ and Al.\ Zamolodchikov 
\cite{Zamolodchikov:1996aa}, based on extensive earlier work by 
many authors (see e.g.\ the reviews \cite{Seiberg:1990eb,%
Teschner:2001rv} for references). Crossing symmetry of the 
conjectured 3-point function was then checked analytically in 
two steps by Ponsot and Teschner \cite{Ponsot:1999uf} and by 
Teschner \cite{Teschner:2001rv,Teschner:2003en}. 
\smallskip 

Different conformally invariant boundary conditions were found 
and studied in several papers\cite{Fateev:2000ik,Teschner:2000md,%
Zamolodchikov:2001ah}. As one might suspect, there exist two 
different types of boundary theories, one describing point-like 
objects, the other corresponding to space-filling branes with 
an additional exponential boundary interaction that is felt by 
the end-points of open strings. It turns out that a point-like 
brane can only sit in the region where the Liouville potential 
$V$ becomes large. The simplest such brane is characterized by 
a 1-point coupling of the form  
\be \label{ZZb}
 B(P) \ = \ \left(\pi \mu \c(b^2)\right)^{-iP/b} \, 
   \frac{2^{7/4}  i \pi P}{\Gamma(1-2iPb^{-1})
    \Gamma(1-2iPb)}    \ \ .       
\ee 
Open strings on this brane possess a discrete spectrum 
without continuous zero modes, in agreement with our  
geometrical intuition (see \cite{Zamolodchikov:2001ah} 
for details). Let us remark that the brane with boundary 
coupling (\ref{ZZb}) is only one in an infinite set which 
is parametrized by two positive integers $(n,m)$. The status
of the boundary conditions with $(n,m) \neq (1,1)$ as physical 
branes in Liouville theory, however, is less clear.  
\smallskip

Extended branes in the Liouville background, on the other 
hand, obviously possess one continuous physical parameter. It 
enters the theory through an additional boundary interaction of 
the form $V_B \sim \mu_B \exp b X$. Note that constant shifts of 
the field $X$ have been used before to rescale the bulk parameter 
$\mu$. Hence, there is no freedom left now to absorb the new
boundary parameter $\mu_B$. Consequently, coupling constants
of the boundary theory display a non-trivial dependence on 
$\mu_B$ as can be seen in the case of the bulk 1-point 
coupling,\cite{Fateev:2000ik}
\be \label{FZZTb} 
 B_s(P) \ = \  \left(\pi \mu \c(b^2)\right)^{-iP/b} \, 
   \cos(2\pi s P) \, \frac{\Gamma(1+2iPb^{-1})
    \Gamma(1+2iPb)} {-2^{5/4} i \pi P}\ \ .
\ee 
Here, the parameter $s$ is related to the boundary coupling 
$\mu_B$ by the relation 
$$ \mu_B^2 \sin \pi b^2\ = \ \mu \cosh^2 \pi s b \ \ . 
$$ 
Each of these extended branes comes with a continuous spectrum 
of open string modes. For imaginary values of the boundary parameter 
$s$, one may also encounter additional 
discrete states\cite{Teschner:2000md}. 
The 3-point vertex for open strings on the extended branes has 
been constructed by Ponsot and Teschner\cite{Ponsot:2001ng}. The 
formulae are more complicated and we refer the interested reader 
to the original papers.

\subsection{Some applications and extensions} 

In this subsection we would like to comment on some of 
the applications and possible extensions of the results 
we outlined above. As we pointed out earlier, Liouville 
theory is a building block for the string theory dual 
of matrix quantum mechanics, 
$$ S_{\rm MQM} \ \sim\ - \beta\int dt \left[ 
     \frac12 (\partial_t M(t))^2 + V(M(t))\right] 
$$ 
where $M(t)$ are hermitian $N\times N$ matrices and $V$ 
is a cubic potential. To be more precise, the duality 
involves taking $N$ and $\beta$ to infinity while keeping 
their ratio $\kappa = N/\beta$ fixed and close to some 
critical value $\kappa_c$. In this double scaling
limit, the matrix model can be mapped to a system of 
non-interacting fermions moving through an inverse 
oscillator potential, one side of which has been 
filled up to a Fermi level at $\Delta \kappa = \kappa - 
\kappa_c \sim g_s^{-1}$. With a quick glance at Figure 8, 
we conclude that the model must be non-perturbatively 
unstable against tunneling of Fermions from the left to 
the right. This instability is reflected in the asymptotic 
expansion of the partition sum and even quantitative 
predictions for the mass $m \sim a/g_s$ of the instantons 
were obtained. The general dependence of brane masses on 
the string coupling $g_s$ along with the specific form of 
the coupling (\ref{ZZb}) have been used recently to 
identify the instanton of matrix quantum mechanics with 
the localized brane in the Liouville model\cite{McGreevy:2003kb,%
Martinec:2003ka,Klebanov:2003km,Alexandrov:2003nn}. In this 
sense, branes had been seen through investigations of matrix 
quantum mechanics more than ten years ago, i.e.\ long 
before their central role for string theory was fully
appreciated. 
\begin{figure} 
\centering
\scalebox{.5}{\epsfbox{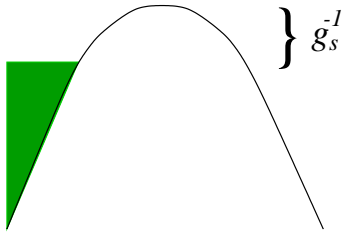}} 
\caption{
In the double scaling limit, the hermitian matrix model can 
be mapped to a system of non-interacting fermions moving 
through an inverse oscillator potential, one side of which 
has been filled up to a Fermi level at $\Delta \kappa = 
\kappa - \kappa_c \sim g_s^{-1}$.}  
\end{figure}
\smallskip 

Concerning the application of Liouville theory to the 
condensation of tachyons, there exist a few quite 
encouraging recent developments. Let us recall that 
the description of tachyon condensation requires
sending the parameter $b$ to $b=i$ and evaluating 
the correlators for imaginary momenta $P$. We can 
at least find one quantity for which this procedure 
may be carried out easily. It is the coupling of closed 
strings to a decaying brane, i.e.\ the 1-point coupling 
$B$ in the theory (\ref{tachact}) with $\Phi =0$ and $\Psi 
= \mu_B$. In fact, from the formula (\ref{FZZTb}) we read 
off that 
$$  \langle\, \exp \left(i E X^0(z,\bz)\right)\, \rangle  
  \ \sim \  \left(\pi \mu\right)^{i E}\, 
   \frac{1}{\sinh \pi E}\ \ .    $$
This result may be checked directly\cite{Sen:2002nu,Larsen:2002wc} 
through perturbative computations\cite{Callan:1994ub,%
Polchinski:my,Recknagel:1998ih}. Other quantities in 
the rolling tachyon background have a much more singular 
behavior at $b=i$. In fact, Barnes' double $\Gamma$-function 
$\Gamma_2(x|b,b^{-1})$ is a well defined analytic function 
as long as ${\rm Re} b \neq 0$. If we send $b \rightarrow i$, 
on the other hand, $\Gamma_2$ diverges. Nevertheless, it can 
be shown\cite{Schomerus:2003vv} that the function $\Upsilon$ 
and the bulk couplings (\ref{CSL}) possess a well defined limit.  
These limits, however, are no longer analytic and hence cannot be 
Wick rotated. It turns out that there are at least two different 
limiting theories. The first one occurs for real momenta 
(Euclidean target space) and it coincides with the interacting 
$c=1$ limit of unitary minimal models that was first discovered 
by Runkel and Watts\cite{Runkel:2001ng}. A limiting bulk theory 
with imaginary momenta (Lorentzian target space) has also been 
constructed\cite{Schomerus:2003vv}. Future investigations will 
show whether this does provide an exact description of closed 
string rolling tachyons.  
\smallskip 

Even though Liouville theory is the only non-rational model that has 
been solved completely, there exist at least partial solutions for
a few other non-compact backgrounds. These include the $H^+_3 \cong$ 
SL($2,\QC$)/SU(2) model, a non-compact relative of the group manifold 
SU(2) and Euclidean version of $AdS_3 \sim $ SL(2,$\QR$). The bulk 
structure constants $C$ of this theory were found several years ago 
by Teschner\cite{Teschner:1997ft} and couplings $B$ of closed strings 
to various branes have been proposed\cite{Ponsot:2001gt,Lee:2001gh}. 
Open string couplings $C$ for non-compact branes in $H^+_3$, on the 
other hand, remain unknown. 

Let us point out that $H_3^3$ model itself suffers from an imaginary 
magnetic field and hence is unphysical. But its solution represented 
a key step toward the construction of two rather important string 
backgrounds. The first concerns the description of strings in $AdS_3$. 
In a series of 
papers\cite{Maldacena:2000hw,Maldacena:2000kv,Maldacena:2001km}, 
amplitudes in this Lorentzian version of the $H^+_3$ model have been 
worked out and interpreted. Another interesting application arises from 
the coset space $H^+_3/ \QR$. This model became famous as the Euclidean 
2D black hole\cite{Witten:1991yr}. It is a non-compact analogue of  
parafermions and hence a close relative of $N=2$ minimal models. 
Non-compact parafermions and $N=2$ minimal models have been argued to 
be T-dual to the Sine-Liouville (as reviewed in\cite{Kazakov:2000pm}) 
and the $N=2$ Liouville theory\cite{Hori:2001ax}, respectively. 
References to the original literature and results on 
branes in these backgrounds can be found in the recent 
literature\cite{Ribault:2003ss,Eguchi:2003ik}.
\bigskip \bigskip

\noindent
{\bf Acknowledgement:} I wish to thank the organizers of the ICMP 2003 
for the opportunity to present this overview.


\newpage
\begin{thebibliography}{99}
\newcommand{\href}[2]{#2} 

\bibitem{Affleck:iv}
I.~Affleck and A.~W.~W.~Ludwig,
{\it Critical Theory Of Overscreened Kondo Fixed Points},
Nucl.\ Phys.\ B {\bf 360} (1991) 641.

\bibitem{AGMOO}
O.~Aharony, S.~S. Gubser, J.~M. Maldacena, H.~Ooguri, and Y.~Oz, {\it Large {N}
  field theories, string theory and gravity}, {\em Phys. Rept.} {\bf 323}
  (2000) 183--386, \href{http://xxx.lanl.gov/abs/hep-th/9905111}{{\tt
  hep-th/9905111}}.

\bibitem{AleSch5}
A.~Yu.~Alekseev and V.~Schomerus, {\it {D-branes} in the {WZW} model}, {\em Phys.
  Rev.} {\bf D60} (1999) 061901,
  \href{http://xxx.lanl.gov/abs/hep-th/9812193}{{\tt hep-th/9812193}}.

\bibitem{AlReSc2}
A.~Yu.~Alekseev, A.~Recknagel, and V.~Schomerus, {\it Non-commutative world-volume
  geometries: {Branes} on {SU(2)} and fuzzy spheres}, {\em JHEP} {\bf 09}
  (1999) 023, \href{http://xxx.lanl.gov/abs/hep-th/9908040}{{\tt
  hep-th/9908040}}.

\bibitem{AlReSc3}
A.~Yu.~Alekseev, A.~Recknagel, and V.~Schomerus, {\it Brane dynamics in background
  fluxes and non-commutative geometry}, {\em JHEP} {\bf 05} (2000) 010,
  \href{http://xxx.lanl.gov/abs/hep-th/0003187}{{\tt hep-th/0003187}}.

\bibitem{AFQS}
A.~Y. Alekseev, S.~Fredenhagen, T.~Quella, and V.~Schomerus, {\it Non-commutative
  gauge theory of twisted {D-branes}},
  \href{http://xxx.lanl.gov/abs/hep-th/0205123}{{\tt hep-th/0205123}}.

\bibitem{Alexandrov:2003nn}
S.~Y.~Alexandrov, V.~A.~Kazakov and D.~Kutasov,
{\it Non-perturbative effects in matrix models and D-branes},
JHEP {\bf 0309} (2003) 057, {\tt hep-th/0306177}.

\bibitem{Barnes1}
E.~Barnes, {\it The theory of the double gamma function},  {\em Philos. Trans.
  Roy. Soc.} {\bf A196} (1901) 265.

\bibitem{BPZ}
A.~A. Belavin, A.~M. Polyakov, and A.~B. Zamolodchikov, {\it Infinite conformal
  symmetry in two-dimensional quantum field theory}, {\em Nucl. Phys.} {\bf
  B241} (1984) 333--380.

\bibitem{Bianchi:1990yu}
M.~Bianchi and A.~Sagnotti,
{\it On The Systematics Of Open String Theories},
Phys.\ Lett.\ B {\bf 247} (1990) 517.

\bibitem{BiFuSc}
L.~Birke, J.~Fuchs, and C.~Schweigert, {\it Symmetry breaking boundary conditions
  and {WZW} orbifolds}, {\em Adv. Theor. Math. Phys.} {\bf 3} (1999) 671--726,
  \href{http://xxx.lanl.gov/abs/hep-th/9905038}{{\tt hep-th/9905038}}.

\bibitem{Callan:1994ub}
C.~G.~.~Callan, I.~R.~Klebanov, A.~W.~W.~Ludwig and J.~M.~Maldacena,
{\it Exact solution of a boundary conformal field theory},
Nucl.\ Phys.\ B {\bf 422}, 417 (1994), {\tt hep-th/9402113}.


\bibitem{Card89}
J.~L. Cardy, {\it Boundary conditions, fusion rules and the {Verlinde} formula},
  {\em Nucl. Phys.} {\bf B324} (1989) 581.

\bibitem{ChuHo1}
C.-S. Chu and P.-M. Ho, {\it Noncommutative open string and {D-brane}}, {\em
  Nucl. Phys.} {\bf B550} (1999) 151,
  \href{http://xxx.lanl.gov/abs/hep-th/9812219}{{\tt hep-th/9812219}}.

\bibitem{Dorn:1994xn}
H.~Dorn and H.~J. Otto, {\it Two and three point functions in {Liouville}
  theory},  {\em Nucl. Phys.} {\bf B429} (1994) 375--388, 
  {{\tt hep-th/9403141}}.

\bibitem{DouHul}
M.~R. Douglas and C.~Hull, {\it D-branes and the noncommutative torus}, {\em
  JHEP} {\bf 02} (1998) 008, \href{http://xxx.lanl.gov/abs/hep-th/9711165}{{\tt
  hep-th/9711165}}.

\bibitem{Douglas:2001ba}
M.~R.~Douglas and N.~A.~Nekrasov,
{\it Noncommutative field theory},
Rev.\ Mod.\ Phys.\  {\bf 73} (2001) 977, {\tt hep-th/0106048}.

\bibitem{Eguchi:2003ik}
T.~Eguchi and Y.~Sugawara,
{\it Modular bootstrap for boundary N = 2 Liouville theory},
JHEP {\bf 0401} (2004) 025, {\tt hep-th/0311141}.


\bibitem{EliSar}
S.~Elitzur and G.~Sarkissian, {\it D-branes on a gauged {WZW} model}, {\em Nucl.
  Phys.} {\bf B625} (2002) 166--178,
  \href{http://xxx.lanl.gov/abs/hep-th/0108142}{{\tt hep-th/0108142}}.

\bibitem{Fateev:2000ik}
V.~Fateev, A.~B. Zamolodchikov and A.~B. Zamolodchikov, {\it Boundary
  {Liouville} field theory. {I: Boundary} state and boundary two-point
  function},  {{\tt hep-th/0001012}}.

\bibitem{FFFS1}
G.~Felder, J.~{Fr\"ohlich}, J.~Fuchs, and C.~Schweigert, {\it The geometry of
  {WZW} branes}, {\em J. Geom. Phys.} {\bf 34} (2000) 162--190,
  \href{http://xxx.lanl.gov/abs/hep-th/9909030}{{\tt hep-th/9909030}}.


\bibitem{Fredenhagen:2000ei}
S.~Fredenhagen and V.~Schomerus,
{\it Branes on group manifolds, gluon condensates, and twisted K-theory},
JHEP {\bf 0104} (2001) 007, {\tt hep-th/0012164}.

\bibitem{FreSch3}
S.~Fredenhagen and V.~Schomerus, {\it D-branes in coset models}, {\em JHEP} {\bf
  02} (2002) 005, \href{http://xxx.lanl.gov/abs/hep-th/0111189}{{\tt
  hep-th/0111189}}.


\bibitem{Fuchs:1999zi}
J.~Fuchs and C.~Schweigert,
{\it Symmetry breaking boundaries. I: General theory},
Nucl.\ Phys.\ B {\bf 558} (1999) 419, 
{\tt hep-th/9902132}.

\bibitem{GabGan}
M.~R. Gaberdiel and T.~Gannon, {\it Boundary states for {WZW} models},
  \href{http://xxx.lanl.gov/abs/hep-th/0202067}{{\tt hep-th/0202067}}.

\bibitem{Gawe2}
K.~Gawedzki, {\it Boundary {WZW}, {G/H}, {G/G} and {CS} theories},
  \href{http://xxx.lanl.gov/abs/hep-th/0108044}{{\tt hep-th/0108044}}.

\bibitem{Gutperle:2002ki}
M.~Gutperle, M.~Headrick, S.~Minwalla and V.~Schomerus,
{\it Space-time energy decreases under world-sheet RG flow},
JHEP {\bf 0301} (2003) 073, {\tt hep-th/0211063}.

\bibitem{Gutperle:2003xf}
M.~Gutperle and A.~Strominger,
{\it Timelike boundary Liouville theory},
Phys.\ Rev.\ D {\bf 67} (2003) 126002, {\tt hep-th/0301038}.

\bibitem{Hori:2001ax}
K.~Hori and A.~Kapustin,
{\it Duality of the fermionic 2d black hole and N = 2 Liouville theory as  mirror
symmetry},
JHEP {\bf 0108} (2001) 045, {\tt hep-th/0104202}.

\bibitem{Kazakov:2000pm}
V.~Kazakov, I.~K.~Kostov and D.~Kutasov,
{\it A matrix model for the two-dimensional black hole},
Nucl.\ Phys.\ B {\bf 622} (2002) 141, {\tt hep-th/0101011}.


\bibitem{Klebanov:1991qa}
I.~R.~Klebanov,
{\it String theory in two-dimensions},
hep-th/9108019.

\bibitem{Klebanov:2003km}
I.~R.~Klebanov, J.~Maldacena and N.~Seiberg,
{\it D-brane decay in two-dimensional string theory},
JHEP {\bf 0307} (2003) 045, {\tt hep-th/0305159}.


\bibitem{Larsen:2002wc}
F.~Larsen, A.~Naqvi and S.~Terashima,
{\it Rolling tachyons and decaying branes},
JHEP {\bf 0302} (2003) 039, {\tt hep-th/0212248}.

\bibitem{Lee:2001gh}
P.~Lee, H.~Ooguri and J.~w.~Park,
{\it Boundary states for AdS(2) branes in AdS(3)},
Nucl.\ Phys.\ B {\bf 632} (2002) 283, {\tt hep-th/0112188}.

\bibitem{Maldacena:1997re}
J.~M.~Maldacena,
{\it The large N limit of superconformal field theories and supergravity},
Adv.\ Theor.\ Math.\ Phys.\  {\bf 2} (1998) 231
[Int.\ J.\ Theor.\ Phys.\  {\bf 38} (1999) 1113], {\tt hep-th/9711200}.

\bibitem{Maldacena:2000hw}
J.~M.~Maldacena and H.~Ooguri,
{\it Strings in AdS(3) and SL(2,R) WZW model. I},
J.\ Math.\ Phys.\  {\bf 42} (2001) 2929,
{\tt hep-th/0001053}.

\bibitem{Maldacena:2000kv}
J.~M.~Maldacena, H.~Ooguri and J.~Son,
{\it Strings in AdS(3) and the SL(2,R) WZW model. II: Euclidean black hole},
J.\ Math.\ Phys.\  {\bf 42} (2001) 2961, {\tt hep-th/0005183}.

\bibitem{Maldacena:2001km}
J.~M.~Maldacena and H.~Ooguri,
{\it Strings in AdS(3) and the SL(2,R) WZW model. III: Correlation  functions},
Phys.\ Rev.\ D {\bf 65} (2002) 106006, {\tt hep-th/0111180}.

\bibitem{MaMoSe1}
J.~Maldacena, G.~W. Moore, and N.~Seiberg, {\it Geometrical interpretation of
  {D-branes} in gauged {WZW} models}, {\em JHEP} {\bf 07} (2001) 046,
  \href{http://xxx.lanl.gov/abs/hep-th/0105038}{{\tt hep-th/0105038}}.

\bibitem{Martinec:2003ka}
E.~J.~Martinec,
{\it The annular report on non-critical string theory},
{\tt hep-th/0305148}.

\bibitem{McGreevy:2003kb}
J.~McGreevy and H.~Verlinde,
{\it Strings from tachyons: The c = 1 matrix reloaded},
JHEP {\bf 0312} (2003) 054
{\tt hep-th/0304224}.

\bibitem{AffOsh2}
M.~Oshikawa and I.~Affleck, {\it Boundary conformal field theory approach to the
  critical two-dimensional {Ising} model with a defect line}, {\em Nucl.
  Phys.} {\bf B495} (1997) 533--582,
  \href{http://xxx.lanl.gov/abs/cond-mat/9612187}{{\tt cond-mat/9612187}}.

\bibitem{PetZub02}
V.~B. Petkova and J.~B. Zuber, {\it Boundary conditions in charge conjugate
  {sl(N)} {WZW} theories}, \href{http://xxx.lanl.gov/abs/hep-th/0201239}{{\tt
  hep-th/0201239}}.

\bibitem{Polchinski:my}
J.~Polchinski and L.~Thorlacius,
{\it Free Fermion Representation Of A Boundary Conformal Field Theory},
Phys.\ Rev.\ D {\bf 50}, 622 (1994), {\tt hep-th/9404008}.

\bibitem{Pollec}
J.~Polchinski, {\it {TASI} lectures on {D}-branes},
  \href{http://xxx.lanl.gov/abs/hep-th/9611050}{{\tt hep-th/9611050}}.

\bibitem{Ponsot:1999uf}
B.~Ponsot and J.~Teschner, {\it Liouville bootstrap via harmonic analysis on a
  noncompact quantum group},  {{\tt
  hep-th/9911110}}.

\bibitem{Ponsot:2001ng}
B.~Ponsot and J.~Teschner, {\it Boundary {Liouville field theory: Boundary}
  three point function},  {\em Nucl. Phys.} {\bf B622} (2002) 309--327,
  {{\tt hep-th/0110244}}.

\bibitem{Ponsot:2001gt}
B.~Ponsot, V.~Schomerus and J.~Teschner,
{\it Branes in the Euclidean AdS(3)},
JHEP {\bf 0202} (2002) 016, {\tt hep-th/0112198}.

\bibitem{Pradisi:1988xd}
G.~Pradisi and A.~Sagnotti,
{\it Open String Orbifolds},
Phys.\ Lett.\ B {\bf 216} (1989) 59.

\bibitem{QueSch1}
T.~Quella and V.~Schomerus, {\it Symmetry breaking boundary states and defect
  lines}, {\em JHEP} {\bf 06} (2002) 028,
  \href{http://xxx.lanl.gov/abs/hep-th/0203161}{{\tt hep-th/0203161}}.

\bibitem{Recknagel:1998ih}
A.~Recknagel and V.~Schomerus,
{\it Boundary deformation theory and moduli spaces of D-branes},
Nucl.\ Phys.\ B {\bf 545}, 233 (1999), {\tt hep-th/9811237}.

\bibitem{Ribault:2003ss}
S.~Ribault and V.~Schomerus,
{\it Branes in the 2-D black hole},
JHEP {\bf 0402} (2004) 019, {\tt hep-th/0310024}.

\bibitem{Runk1}
I.~Runkel, {\it Boundary structure constants for the {A-series} {Virasoro} minimal
  models}, {\em Nucl. Phys.} {\bf B549} (1999) 563,
  \href{http://xxx.lanl.gov/abs/hep-th/9811178}{{\tt hep-th/9811178}}.

\bibitem{Runkel:2001ng}
I.~Runkel and G.~M.~T. Watts, {\it A non-rational {CFT} with c = 1 as a limit
  of minimal models},  {\em JHEP} {\bf 09} (2001) 006, {{\tt hep-th/0107118}}.

\bibitem{Scho}
V.~Schomerus, {\it D-branes and deformation quantization}, MLI-13-1998-99, 
  {\em JHEP} {\bf 06}
  (1999) 030, \href{http://xxx.lanl.gov/abs/hep-th/9903205}{{\tt
  hep-th/9903205}}.

\bibitem{Schomerus:2002dc}
V.~Schomerus,
{\it Lectures on branes in curved backgrounds},
Class.\ Quant.\ Grav.\  {\bf 19} (2002) 5781, {\tt hep-th/0209241}.


\bibitem{Schomerus:2003vv}
V.~Schomerus,
{\it Rolling tachyons from Liouville theory},
JHEP {\bf 0311} (2003) 043, {\tt hep-th/0306026}.

\bibitem{Seiberg:1990eb}
N.~Seiberg,
{\it Notes On Quantum Liouville Theory And Quantum Gravity},
Prog.\ Theor.\ Phys.\ Suppl.\  {\bf 102} (1990) 319.

\bibitem{SeiWit99}
N.~Seiberg and E.~Witten, {\it String theory and noncommutative geometry}, 
{\em JHEP} {\bf 09} (1999) 032, 
\href{http://xxx.lanl.gov/abs/hep-th/9908142}{{\tt hep-th/9908142}}.

\bibitem{Sen:2002nu}
A.~Sen,
{\it Rolling tachyon},
JHEP {\bf 0204} (2002) 048, 
{\tt hep-th/0203211}.

\bibitem{Teschner:1997ft}
J.~Teschner,
{\it On structure constants and fusion rules in the SL(2,C)/SU(2) WZNW  model},
Nucl.\ Phys.\ B {\bf 546} (1999) 390, {\tt hep-th/9712256}.


\bibitem{Teschner:2000md}
J.~Teschner, {\it Remarks on {Liouville} theory with boundary},
  {{\tt hep-th/0009138}}.

\bibitem{Teschner:2001rv}
J.~Teschner, {\it Liouville theory revisited},  {\em Class. Quant. Grav.} {\bf
  18} (2001) R153--R222, {{\tt
  hep-th/0104158}}.

\bibitem{Teschner:2003en}
J.~Teschner, {\it A lecture on the {Liouville} vertex operators},
 {{\tt hep-th/0303150}}.

\bibitem{Witten:1991yr}
E.~Witten,
{\it On string theory and black holes},
Phys.\ Rev.\ D {\bf 44} (1991) 314.

\bibitem{Zamolodchikov:1996aa}
A.~B. Zamolodchikov and A.~B. Zamolodchikov, {\it Structure constants and
  conformal bootstrap in {Liouville} field theory},  {\em Nucl. Phys.} {\bf
  B477} (1996) 577--605, {{\tt hep-th/9506136}}.

\bibitem{Zamolodchikov:2001ah}
A.~B. Zamolodchikov and A.~B. Zamolodchikov, {\it Liouville field theory on a
  pseudosphere},  {{\tt
  hep-th/0101152}}.

\end{thebibliography}
\end{document}